# Statistical Study of Deep Sub-Micron Dual-Gated Field-Effect Transistors on Monolayer CVD Molybdenum Disulfide Films

Han Liu,[†] Mengwei Si,[†] Sina Najmaei,[‡] Adam T. Neal,[†] Yuchen Du,[†] Pulickel M. Ajayan,[‡] Jun Lou[‡] and Peide D. Ye[†,*]

[†] School of Electrical and Computer Engineering and Birck Nanotechnology Center, Purdue University, West Lafayette, IN 47907, USA

[‡] Department of Mechanical Engineering and Materials Science, Rice University, Houston, TX 77005, USA

*Correspondance to: yep@purdue.edu

# Abstract

Monolayer Molybdenum Disulfide ($MoS_2$) with a direct band gap of 1.8 eV is a promising two-dimensional material with a potential to surpass graphene in next generation nanoelectronic applications. In this letter, we synthesize monolayer $MoS_2$ on $Si/SiO_2$ substrate via chemical vapor deposition (CVD) method and comprehensively study the device performance based on dual-gated $MoS_2$ field-effect transistors. Over 100 devices are studied to obtain a statistical description of device performance in CVD $MoS_2$. We examine and scale down the channel length of the transistors to 100 nm and achieve record high drain current of 62.5 mA/mm in CVD monolayer $MoS_2$ film ever reported. We further extract the intrinsic contact resistance of low work function metal Ti on monolayer CVD $MoS_2$ with an expectation value of 175 Ω·mm, which can be significantly decreased to 10 Ω·mm by appropriate gating. Finally, field-effect mobilities ($\mu_{FE}$) of the carriers at various channel lengths are obtained. By taking the impact of contact resistance into account, an average and maximum intrinsic $\mu_{FE}$ is estimated to be 13.0 and 21.6 $cm^2/Vs$ in monolayer CVD $MoS_2$ films, respectively.

Two-dimensional (2D) layered crystals have attracted great attention since the advent of graphene.[1-6] These materials are made of individual layers bonded by van der Waals forces and can be easily exfoliated to obtain atomically thin crystals.[7] Despite its ultra-high charge carrier mobility due to its linear energy dispersion in momentum space, graphene's gapless electronic structure restrains its applications in digital circuits.[8] In comparison, some members of transition metal dichalcogenides (TMDs), another family of layered materials, provide semiconducting substitutes to graphene. Bulk $MoS_2$ with an indirect band gap of 1.2 eV and a direct band gap of 1.8 eV in monolayers is a promising member of this group.[9] Applications such as field-effect transistors, chemical sensors, photonic detectors, and integrated circuits have been explored.[10-14] These studies were mostly done on single or multi-layer $MoS_2$ flakes peeled from bulk via scotch tape technique.[7] Though mechanical exfoliation offers an easy and feasible way to obtain $MoS_2$ thin flakes with various thicknesses for fundamental research, the low yield and non-uniformity limits its practical applications.

Chemical vapor deposition (CVD) has been proved to be a good method for 2D crystal growth. Typically, this synthetic route provides a low cost path to high-quality, large-area and thin films. A variety of 2D crystals, such as graphene, boron nitride, topological insulators, and $MoS_2$ have been successfully synthesized by CVD methods.[15-17] The early attempts to obtain large area $MoS_2$ have relied on the solid state sulfurization of molybdenum and molybdenum compounds, such as $(NH_4)_2MoS_4$.[18-19] However, they suffer from non-uniformity in thickness, small grain-sizes, and difficulty in precursor preparation. In addition, the reported lower charge carrier mobility in these CVD films is not favorable for device applications. Additionally, the sulfurization of $MoO_3$ has been comprehensively studied and is a main approach in $MoS_2$ synthesis, as $MoO_3$ has a lower melting and evaporation temperature. Several studies have shown the feasibility of CVD $MoS_2$ synthesis from sulfurization of $MoO_3$ on $Si/SiO_2$ or sapphire.[20-21] However, the electrical properties of these CVD films have not been comprehensively studied. In this work, highly-crystalline monolayer $MoS_2$ crystals were obtained from a CVD based procedure, on which we fabricated over 100 transistors with channel lengths scaled down to deep sub-micron region. Several intrinsic device parameters from each transistor are extracted, including the drain current, contact resistance and carrier mobility. By statistically studying the devices performance, we are able to achieve a comprehensive understanding of the transistor behavior based on monolayer CVD $MoS_2$ films.

The synthesis of $MoS_2$ thin films was carried out in a vapor phase deposition process. As shown in Figure 1(a), the precursors, $MoO_3$ nanoribbons and sublimated sulfur, were placed separately in a quartz tube. $MoO_3$-covered silicon substrates, along with several clean substrates designated for the growth of $MoS_2$ were placed close to each other at the center of the furnace, flushed with nitrogen at a constant flow of 200 sccm. Clean heavily doped silicon substrates coated with 285 nm of $SiO_2$ was used, which also facilitates the direct device fabrication. A container with 0.8 - 1.2 grams of sublimated sulfur was placed at a location reaching an approximate maximum temperature of 600 ºC at the opening of the furnace. The center of the furnace was gradually heated from room temperature to 550 ºC at a ramping rate of ~20 ºC/min. At 550 ºC, the sulfur slowly evaporated and the chamber was then heated to 850 °C at a slower ramping rate of ~5 °C/min. The temperature of chamber was then maintained at this temperature for 10-15 min, and then the chamber was naturally cooled back to room temperature. The chamber pressure and the closely related sulfur concentration in the chamber were monitored using the guidelines described in reference 21 to optimize the growth of large triangular single crystals. These materials are convenient for device fabrication and allow us to avoid complexities in transport caused by grain boundaries. A more detailed description of the synthesis process and growth kinetics can be found in our previous work.[21] Figure 1(b) represents an image of these single crystalline monolayer $MoS_2$ samples under optical microscope. Typical triangular domain side lengths vary between 10 to 20 microns. The atomic force microscope (AFM) image of the CVD $MoS_2$ is shown in Figure 1(c). The thickness of the flake is measured to be 0.85 nm, showing monolayer $MoS_2$ with a good thickness uniformity.

The monolayer CVD $MoS_2$ samples were then used for device fabrication. To start with, source/drain regions were defined by e-beam lithography. The contact width of source and drain was 2 μm. The source and drain spacing, or the channel lengths, were chosen as 100, 200, 500 nm, and 1 μm so that they cover the ranges from long channel to short channel. Ti/Au of 20/60 nm was then deposited by e-beam evaporator as the metal contacts. After the lift-off process, a 0.8-1 nm Al seeding layer was deposited at the rate of 0.1 Å/s on the whole sample to facilitate dielectric growth.[22-24] The samples were then aged overnight in atmosphere to secure a complete oxidation of ~1 nm $Al_2O_3$ as the seeding layer. After that, a 15 nm $Al_2O_3$ was deposited by atomic layer deposition (ALD) at 200 °C using trimethylaluminum (TMA) and water. Pulse times were 0.8 and 1.2 seconds for TMA and water, and purge times were 6 and 8 seconds, respectively. Previously, we had a temperature dependent study on direct ALD growth on 2D crystals.[22] However, we noticed that the insertion of the seeding layer would significantly enhance the yield of devices and minimize the leakage currents. The AFM images of $Al_2O_3$ growth on monolayer CVD $MoS_2$ crystals are shown in Figure S1. Finally, the top gate regions were defined by e-beam lithography again. The length of the top gate was chosen to be similar to the channel length, i.e. $L_g=L_{ch}$, which reduces the access resistance of the top-gated devices. E-beam evaporated Ni/Au of 20/60 nm was deposited as the top gate metal. A total of 120 devices were fabricated and a statistical study of their transport properties was performed using a Keithley 4200 Semiconductor Characterization Systems.

Figure 2(a) shows a schematic configuration of the device structures and the dual-gated $MoS_2$ field-effect transistors. Heavily doped silicon and 285 nm $SiO_2$ serve as the back gate and gate dielectric, while Ni/Au and 16 nm $Al_2O_3$ serve as the top gate and gate dielectric materials. By using the dual-gate structure, carrier density in the channel can be modulated by either gate. Figure 2(b) shows typical output characteristics of a 100 nm channel length device from back-gate modulation. The top-gate is grounded during the measurement to eliminate the capacitance coupling effect.[25] The linear current-voltage relationship at low drain bias indicates good "ohmic" contact at source/drain regions. At 2 V drain voltage, we achieve the highest drain current of 62.5 mA/mm at 100V back gate voltage, which is equivalent to a vertical field of 3.5 MV/cm. To our best knowledge, this is the highest drain current for CVD $MoS_2$ based transistors ever reported.[26] However, suffering from the thick back gate dielectric and a large interface trap density (~$1.6 \times 10^{13}$ /cm$^2$·eV) at the SiO$_2$/MoS$_2$ interface, the transconductance is only 0.83 mS/mm. A full scale characterization and the estimation of interface trap density are described in the supporting information and Figure S2. Both of the drain current and transconductance can be further improved by optimizing the synthesis process and device fabrication. We also notice the big difference between top-gate and back-gate modulations in the same dual-gate device here. The family of output curves of a top-gated device is shown in Figure 2(c). During this measurement, the back-gate is grounded.[25] Channel length of this device is 1 μm. Long-channel device is compared here where the device is operated in the diffusive regime and carrier velocity saturation can be ignored so that we can neglect the non-ideal factors. We achieve the highest on-current of 2.71 mA/mm only for the top-gate modulation where top gate bias and drain bias are 2 V. The highest drain current of this device under the same 2 V drain bias at 100 V back-gate bias is 14.9 mA/mm, 5 times larger than the drain current from top-gating. Since the channel region is fully gated by either top or back gates, this current difference is mostly originated from the variance in contact resistance. For back-gated devices, the carrier density in MoS$_2$ under source/drain metal contacts will be increased at higher positive gate bias. Since the source/drain regions are not heavily doped, as in conventional semiconductors, the contact resistance is mainly determined by the effective Schottky barrier height at metal/semiconductor interface.[27-28] With large gate bias, the conduction band bends downward at the metal/semiconductor interface to enhance tunneling current thus reduce contact resistance. Therefore it facilitates carrier injection from metal contacts to MoS$_2$. It also can be easily understood by electrostatic doping of MoS$_2$ underneath source/drain contacts to reduce contact resistance with positive back-gate bias. On the contrary, the top gate has no effect on the carrier density of $MoS_2$ under the source/drain thus the contact resistance remains constant even at large positive bias. The contact resistance issue would be further discussed in later parts.

Compared to molecular beam epitaxial (MBE) grown crystals, devices made on CVD samples are usually with larger performance variance. Usually, CVD samples have small domain size, random crystal orientation, the existence of grain boundaries and large defect density. These non-ideal phenomena have been observed in both graphene and $MoS_2$ CVD films.[21,29] Therefore, a statistical study of the key parameters of devices is necessary in order to gain a comprehensive understanding of the electrical properties of the CVD grown samples. Figure 3 (a) to (d) show the distributions of the maximum drain currents at different channel lengths. All values are extracted at 2 V drain bias and back-gate voltage of 100 V. The variance in device threshold voltages is ignored in the context of this statistical study. 17-27 devices are studied for each channel length. As expected, these figures show a broad distribution for the maximum drain current at all channel lengths. This can be attributed to the non-uniformity of material synthesis and device fabrication; however they show a normal distribution in all sets of data. The average and standard deviation of the measured drain currents are 36.7±14.2, 27.1±12.2, 22.3±10.0 and 12.9±5.0 mA/mm for 100, 200, 500 nm and 1 μm channel lengths, respectively. The channel-length dependent average value with standard deviation and the maximum value measured are plotted in Figure 4. From long channel to short channel devices, the increase in drain current indicates the scaling properties in CVD $MoS_2$ transistors. If we assume the carriers follow the diffusive transport in all channel lengths, the saturated drain current exhibits $I_{ds,sat} = \frac{1}{2} \mu_{FE} C_{ox} \frac{W}{L} (V_{gs} - V_{th})^2$ by Square Law Theory, where $\mu_{FE}$ is the field-effect mobility, $C_{ox}$ is the gate oxide capacitance, $W$ and $L$ are the width and length of the channel, $V_{gs}$ and $V_{th}$ are the gate bias and threshold voltage.[30] This shows that drain current is inversely proportional to the channel length, and this trend is drawn by the red dashed line in Figure 5. The deviation of drain current at shorter channel lengths is associated with two factors. One is that contact resistance does not scale with channel length, and the impact of this factor is magnified in transistors with ultrathin body semiconductor due to the larger contact resistance.[27] Another reason is the velocity saturation in CVD MoS$_2$ transistors, where carriers have been approaching the maximum drift velocity in short channel devices, as observed in most conventional semiconductor transistors.

As we have stated before, the major issue for 2D semiconductor based transistors is the existence of a large contact resistance ($R_c$), which drastically restrains the drain current.[27] The fundamental reason for the large $R_c$ is that the Fermi level pinning at the metal/semiconductor interface that results in a notable Schottky barrier height. However, their atomically thin body makes it difficult for the realization of source/drain engineering such as ion implantation, as a common approach to dope the source/drain regions and reduce contact resistance in traditional semiconductor devices. Thus it's important to study the contact properties, especially in transistors based on monolayer MoS$_2$. Here we develop a simple and effective method to extract the intrinsic contact resistance, or the contact resistance without electrostatic doping, in CVD MoS$_2$ transistors. The total resistance of the transistor, $R_{tot}$, is the sum of the contact resistance and channel resistance, i.e. $R_{tot} = 2R_c + R_{ch}$. Here $2R_c$ represents the contact resistance of both source and drain leads. For each dual-gated device, the on-current from back-gate modulation and top-gate modulation is very different, even they cover the same area of channel region. This is mostly due to the difference in contact resistance. The global back gate is able to modulate the carrier density of $MoS_2$ under source/drain metal contacts, thus reduce the contact resistance by electrostatic doping. However, this effect from top gate is screened by source/drain contact metals, thus the carrier density in $MoS_2$ at source/drain regions, and hence the contact resistance, remain constant during top-gate sweeping. The observed current ratio between back-gate and top-gate modulation is around 10 (See Figure S3). This means that for top-gate modulation, the contact resistance is over 10 times larger than the channel resistance, i.e. $2R_c \gg R_{ch}$. Statistical values and detailed calculations are provided in supporting information and Figure S3. This clearly explains a quick saturation in transfer curves of top-gated devices shown in the inset of Figure 5. In top-gated devices, $R_c$ is fixed at a large and constant value, once the $R_{ch}$ is much smaller than $2R_c$ during top-gate sweep, $R_{tot}$ doesn't change much anymore even if we increase the top-gate bias. Beyond this point, we can assume $R_{tot} \approx 2R_c$. We extract the total resistance in all top-gated devices and plot the distributions in Figure 5. We achieve the expectation value of $2R_c$ to be about 350 Ω·mm at zero back-gate bias or without electrostatic doping from back-gate. This huge number is almost two to three orders larger than the desired value in conventional semiconductor devices. Besides the Schottky barrier issue stated before, the single layer nature also has a remarkable contribution to the large contact resistance. Previous study about metal contacts on graphene has revealed that the contact resistivity ($\rho_c$) is determined by contact width instead of the contact area, i.e. $\rho_c = R_c W$.[31] This means that in the case of graphene, the current flows mainly along the edge of the graphene/metal contact. In other words, the current crowding takes place at the edge of the contact metal. We believe this also applies to monolayer MoS2 since it's more resistive with lower mobility than graphene. The contact resistance $R_c$ could be reduced to 10 Ω·mm at 100 V back-gate bias, which would be discussed in later parts. A comprehensive study on the metal/monolayer MoS2 contact is required in the future to elucidate these issues.

Next, we study the carrier mobility in monolayer CVD MoS2 field-effect transistors. In classical theories, the carrier mobility in the bulk and at the interface is different, thus the field-effect mobility is dependent on the gate voltage, which can be written as $\mu_{FE} = \dfrac{\mu_{FE}^0}{1+\theta(V_{gs}-V_{th})}$, where $\mu_{FE}^0$ and $\theta$ are two constants. Since increasing the gate bias would push the carriers to the boundary of semiconductor and gate oxide, thus the carrier mobility would be lowered due to surface scattering. However, in our case the channel material is only atomic layer thick and this effect would have a much smaller impact, hence we assume the mobility is nearly a constant value during gate sweep or $\theta = 0$. As a common approach, the field-effect mobility is extracted from the maximum transconductance on the transfer characteristics, which follows $g_m = \dfrac{\partial I_{ds}}{\partial V_{gs}} = \mu_{FE} C_{ox} \dfrac{W}{L} V_{ds}$.

Field-effect mobility at various lengths is calculated and the distributions are plotted in Figure 6 (a) to (d). We achieve the average values of extrinsic field-effect mobility to be 2.67±0.91, 4.58±1.71, 8.52±2.97 and 10.52±3.41 cm$^2$/V·s for 100, 200, 500 nm and 1 μm channel lengths, respectively. The decrease in mobility at shorter channel lengths is mostly because we assume the drift velocity $\upsilon_d$ increases linearly with drain bias by applying $v_d = -\mu E$ this equation. However, due to velocity saturation, the product of carrier mobility and lateral electric field is fixed at a constant number. The electric field in the channel increases with channel length down scaling, thus the *calculated* mobility decreases. This is consistent with our observation of drain current saturation even when we push the channel length to smaller values.

Finally, we estimate the intrinsic carrier mobility in monolayer CVD MoS$_2$ transistors by subtracting the influence of the significant contact resistance. Due to the large contact resistance we have stated, the intrinsic mobility could be very different from the extrinsic values. Therefore, it is necessary to give a reasonable estimation of the contact resistance under gate bias and correctly estimate the intrinsic field-effect mobility. The back-gated maximum transconductance achieved in our measurement range is at 100 V gate bias. We assume there is no velocity saturation in devices channel length of 500 nm or above so as to ignore the difference in mobility between 500 nm and 1μm channel length devices. For back-gated devices, the contact resistance is a function of gate bias, and the channel resistance is a function of both gate bias and channel length, i.e. $R_{tot} = 2R_c(V_{gs}) + R_{ch}(V_g, L)$. Since $R_{ch}(V_{gs}, L)|_{L=1\mu m} \approx 2R_{ch}(V_{gs}, L)|_{L=500nm}$ at fixed V$_{gs}$, therefore R$_c$ can be extracted at a certain V$_{gs}$. A detailed statistical estimation of the R$_c$ is provided in the supporting information. By applying this method, we get an expectation value of 2R$_c$ to be 20 Ω·mm. Therefore, the intrinsic carrier mobility is estimated by dividing the ratio between total resistances over channel resistance, i.e. $\mu' = \mu \left(\dfrac{R_{ch}}{R_{tot}}\right)^{-1} = \mu \left(1 - \dfrac{2R_c}{R_{tot}}\right)^{-1}$. We get averaged

values of 6.11±2.07, 7.72±2.89, 12.59±4.38 and 13.02±4.22 cm$^2$/V·s for 100, 200, 500 nm and 1 μm channel length and a maximum value of ~21.6 cm$^2$/V·s at long channel regions of the CVD monolayer MoS$_2$ transistors, as plotted in Figure 7. These mobility values provide the low limit for our CVD MoS$_2$ films. The significant interface trap density, which degrades the measured $g_m$, has not been considered in this estimation.

In summary, we have synthesized monolayer MoS$_2$ films by CVD method and have statistically studied their electrical properties. Devices with channel length down to deep sub-micron region were fabricated. We achieve a maximum drain current of 62.5 mA/mm at 2 V drain bias for 100 nm channel length device. We also revealed the existence of large contact resistance for metal contacts on CVD monolayer MoS$_2$ films up to R$_c$=175 Ω·mm, which could be reduced to 10 Ω·mm under 3.5MV/cm vertical field. We also extract the field-effect mobility and acquire its intrinsic values by subtracting the contact resistance. The maximum value of intrinsic field effect mobility in CVD monolayer MoS$_2$ is calculated to be 21.6 cm$^2$/V·s. We demonstrate the importance to understand metal/MoS$_2$ interface and significantly reduce the contact resistance for the real device applications of CVD MoS$_2$ films.

**Supporting Information**

AFM images of ALD integration on monolayer CVD MoS$_2$ with and without Al seeding layers (Figure S1), device performance from top-gate modulation and back-gate modulation on the same device (Figure S2), method of estimating intrinsic contact resistance and distributions (Figure S3), method of estimating contact resistance under gate bias (Figure S4), and method for mobility correction (Figure S5) are provided in supporting information.

**Reference**

1. Novoselov, K. S.; Geim, A. K.; Morozov, S. V.; Jiang, D.; Katsnelson, M. I.; Grigorieva, I. V.; Dubonos, S. V.; Firsov, A. A. *Nature* **2005**, 438, 197.
2. Zhang, Y. B.; Tan, Y. W.; Stormer, H. L.; Kim, P. *Nature* **2005**, 438, 201.
3. Chen, Y. L.; Analytis, J. G.; Chu, J. H.; Liu, Z. K.; Mo, S. K.; Qi, X. L.; Zhang, H. J.; Lu, D. H.; Dai, X.; Fang, Z.; Zhang, S. C.; Fisher, I. R.; Hussain, Z., Shen, Z.X. *Science* **2009**, 325, 178.
4. Fu, L.; Kane, C. L.; Mele, E. J. *Phys. Rev. Lett.* **2005**, 98, 106803.
5. Radisavljevic, B.; Radenovic, A.; Brivio, J.; Giacometti, V.; Kis, A. *Nat. Nanotechnol.* **2011**, 6, 147.
6. Fang, H.; Chuang, S.; Chang, T. C.; Takei, K.; Takahashi, T.; Javey, A. *Nano Lett.* **2012**, 12, 3788.
7. Novoselov, K. S.; Jiang, D.; Schedin, F.; Booth, T. J.; Khotkevich, V. V.; Morozov, S. V.; Geim, A. K. *Proc. Natl. Acad. Sci. USA* **2005**, 102, 10451.
8. Wu, Y. Q.; Lin, Y.-M.; Bol, A. A.; Jenkins, K. A.; Xia, F. N.; Farmer, D. B.; Zhu, Y.; Avouris, P. *Nature* **2011**, 472, 74.
9. Mak, K. F.; Lee, C.; Hone, J.; Shan, J.; Heinz, T. F. *Phys. Rev. Lett.* **2010**, 105, 136805.
10. Liu, H.; Ye, P. D. *IEEE Elect. Dev. Lett.* **2012**, 33, 546.11. Liu, H., Gu, J.J., Ye, P. D. *IEEE Elect. Dev. Lett.* **2012**, 33, 1273.
12. Wang, H.; Yu, L.; Lee, Y.-H.; Shi, Y.; Hsu, A.; Chin, M. L.; Li, L.-J.; Dubey, M.; Kong, J.; Palacios, T. *Nano Lett.* **2012**, 12, 4674.
13. Perkins, F. K.; Friedman, A. L.; Cobas, E.; Campbell, P. M.; Jernigan, G. G. and Jonker, B. T. *Nano Lett.* **2013**, 13, 668.
14. Buscema, M.; Barkelid, M.; Zwiller, V.; van der Zant, H.; Steele, G.A.; Andres Castellanos-Gomez, A. *Nano Lett.* **2013**, 13, 358.
15. Li, X. S.; Cai, W. W.; An, J. H.; Kim, S.; Nah, J.; Yang, D. X.; Piner, R. D.; Velamakanni, A.; Jung, I.; Tutuc, E.; Banerjee, S. K.; Colombo, L.; Ruoff, R. S. *Science* **2009**, 324, 1312.
16. Song, L.; Ci, L.; Lu, H.; Sorokin, P. B.; Jin, C.; Ni, J.; Kvashnin, A. G.; Kvashnin, D. G.; Lou, J.; Yakobson, B. I. *Nano Lett.* **2010**, 10, 3209.
17. Kong, D.; Dang, W.; Cha, J. J.; Li, H.; Meister, S.; Peng, H.; Liu, Z.; Cui, Y. *Nano Lett.* **2010**, 10, 2245.
18. Liu, K.-K.; Zhang, W.; Lee, Y.-H.; Lin, Y.-C.; Chang, M.-T.; Su, C.-Y.; Chang, C.-S.; Li, H.; Shi, Y.; Zhang, H.; Lai, C.-S; Li, L.-J. *Nano Lett.* **2012**, 12, 1538.

19 Zhan, Y.; Liu, Z.; Najmaei, S.; Ajayan, P. M.; Lou, J. *Small* **2012**, 8, 966.
20. Lee, Y.-H.; Zhang, X.-Q.; Zhang, W.; Chang, M.-T.; Lin, C.-T.; Chang, K.-D.; Yu, Y.-C.; Wang, J. T.-W.; Chang, C.-S.; Li, L.-J.; Lin, T.-W. *Adv. Mater.* **2012**, 24, 2320.
21. Najmaei, S.; Liu, Z; Zhou, W.; Zou, X.; Shi G.; Lei, S.; Yakobson, B. I.; Idrobo, J.-C.; Ajayan, P. M.; Lou, J. **2013**, arXiv:1301.2812.
22. Liu, H.; Xu, K.; Zhang, X.; Ye, P. D. *Appl. Phys. Lett.* **2012**, 100, 152115.
23. Kim, S.; Nah, J.; Jo, I.; Shahrjerdi, D.; Colombo, L.; Yao, Z.; Tutuc, E.; Banerjee, S. K. *Appl. Phys. Lett.* **2009**, 94, 062107.
24. Shen, T.; Gu, J. J.; Xu, M.; Wu, Y. Q.; Bolen, M. L.; Capano, M. A.; Engel, L. W.; Ye, P. D. *Appl. Phys. Lett.* **2009**, 95, 172105.
25. Fuhrer, M. S. and Hone, J. **2013**, arXiv:1301.4288.
26. Wang, H. ; Yu, L.; Lee, Y.-H.; Fang, W.; Hsu, A.; Herring, P.; Chin, M.; Dubey, M.; Li, L.-J.; Kong, J.; Palacios. T. *IEEE IEDM Technical Digest* **2012**, 88.
27. Liu, H.; Neal, A. T.; Ye, P. D. *ACS Nano* **2012**, 6, 8563.
28. Das, S.; Chen, H.-Y.; Penumatcha, A. V.; Appenzeller, J. *Nano Lett.* **2013**, 13, 100.
29. Huang, P. Y.; Ruiz-Vargas, C. S.; van der Zande, A. M.; Whitney, W. S.; Levendorf, M.; Kevek, J.; Garg, S.; Alden, J. S.; Hustedt, C. J.; Zhu, Y.; Park, J.; McEuen, P. L.; Muller, D. A. *Nature* **2011**, 469, 389.
30. Sze, S. M.; Ng, K. K. *Physics of semiconductor devices*, Wiley Interscience: Hoboken, NJ, **2007**.
31. Nagashio, K.; Nishimura, T.; Kita, K.; Toriumi, A. *Appl. Phys. Lett.* **2010**, 97, 143514.

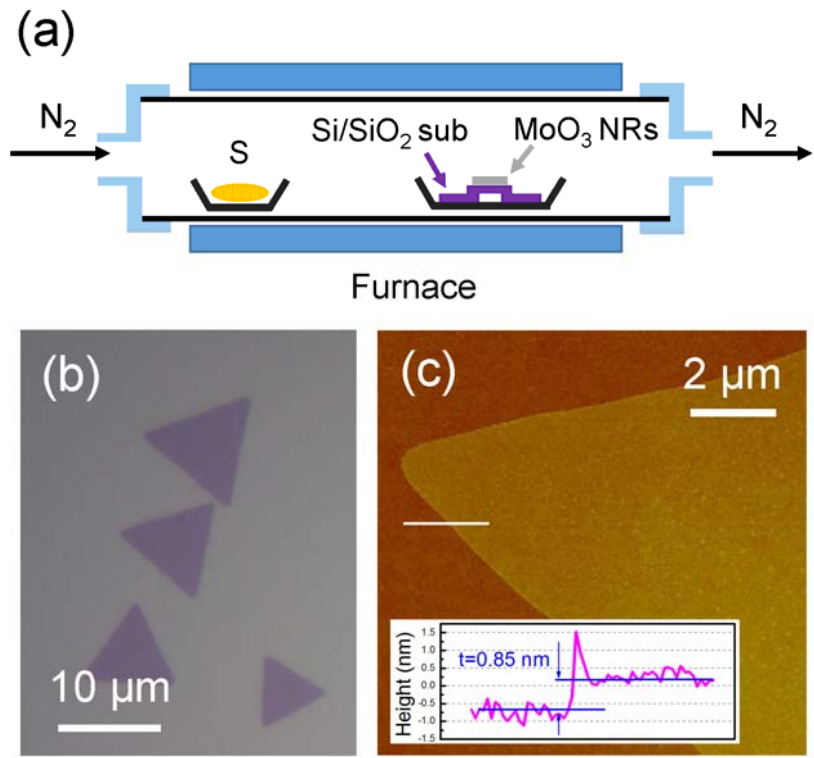

Figure 1: (a) Schematic view of CVD synthesis of monolayer MoS$_2$ in a furnace. (b) Optical micrograph of single crystalline monolayer CVD MoS$_2$. Scale bar is 10 μm. (c) AFM image of a synthesized MoS$_2$ crystal showing good single layer thickness uniformity. Scale bar is 2 μm.

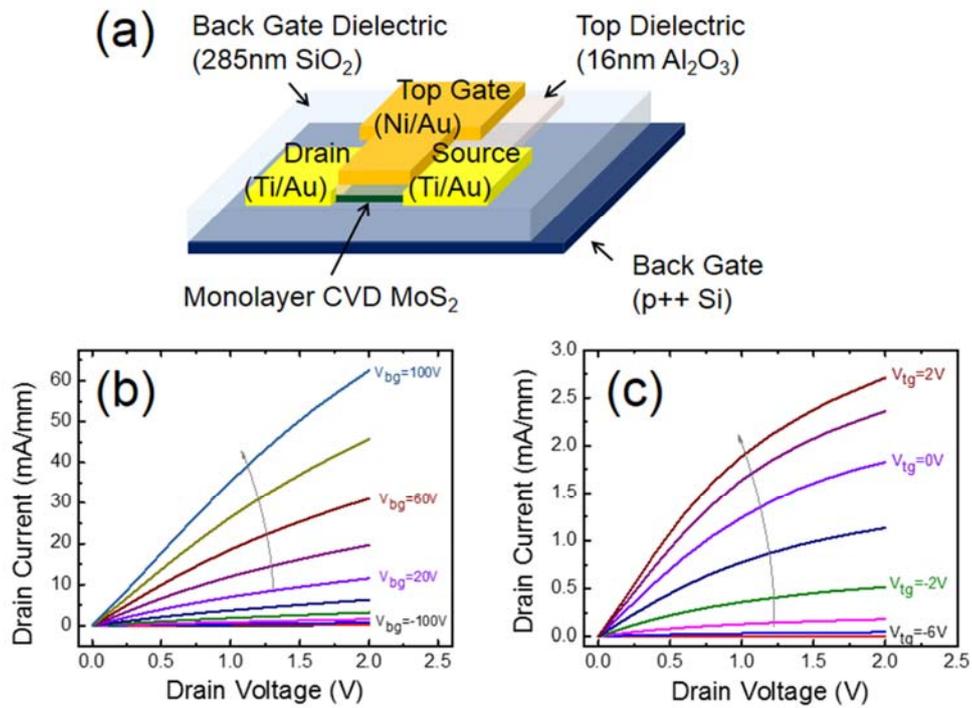

Figure 2: (a) Schematic view of a dual-gate field-effect transistor based on monolayer CVD MoS2. Heavily doped Si and 285 nm SiO2 are used as back-gate and dielectric; Ni/Au and 16 nm Al2O3 are used as top gate and dielectric. Ti/Au are the contacts for both source and drain. (b) Output curves for a 100 nm channel length device under different back-gate bias showing maximum drain current to be over 60 mA/mm. (c) Output curves of a 1 μm channel length device under top-gate bias.

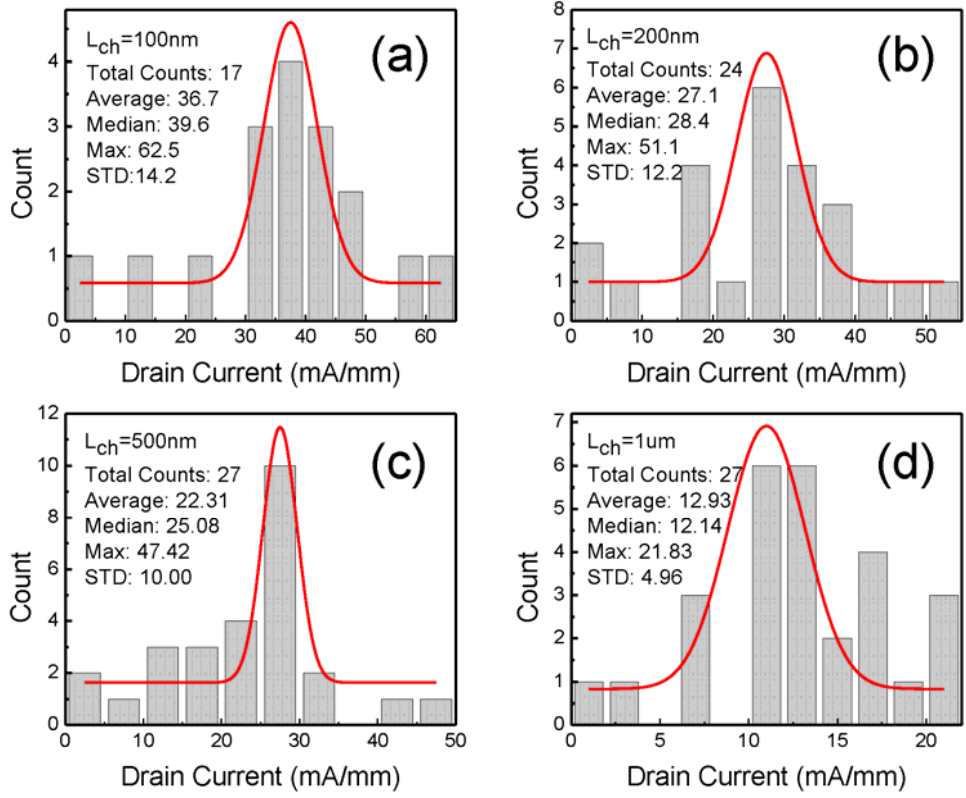

Figure 3: Maximum drain current distributions of (a) 100 nm, (b) 200 nm, (c) 500 nm and (d) 1 μm devices extracted from back-gate modulation. Average values of 36.7±14.2, 27.1±12.2, 22.3±10.0 and 12.9±5.0 mA/mm for 100, 200, 500 nm and 1μm channel lengths are acquired, respectively.

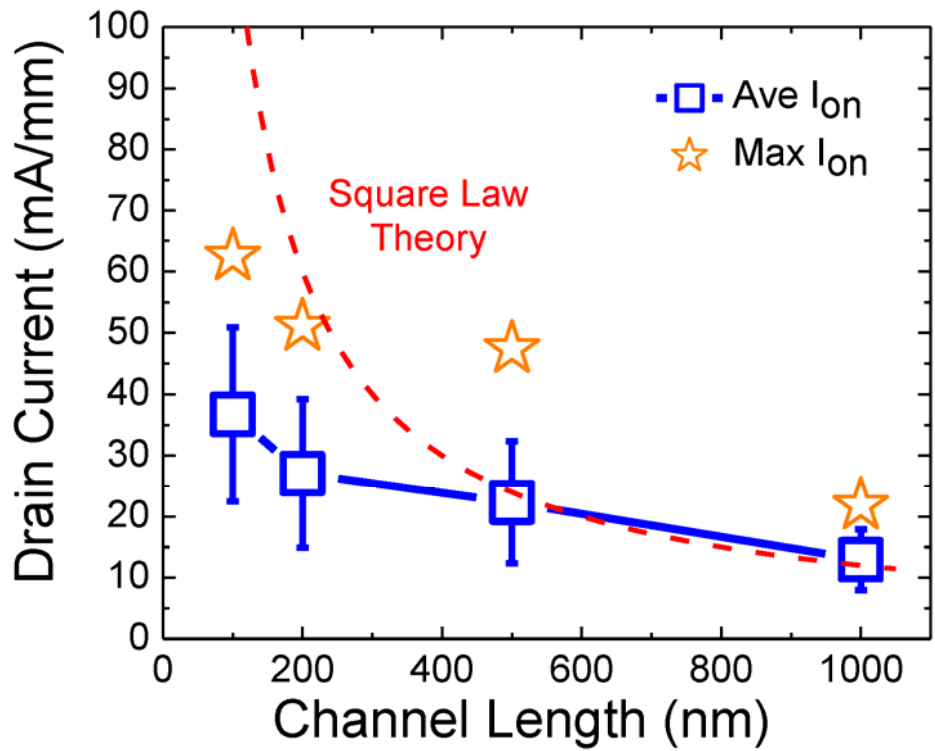

Figure 4: Averaged values of maximum drain current at all channel lengths. Maximum values in measurement are also plotted. Red dashed line shows Square Law Model prediction of channel length dependent drain current which fits long channel device performance. The deviation at short channel regions is due to contact resistance and velocity saturation.

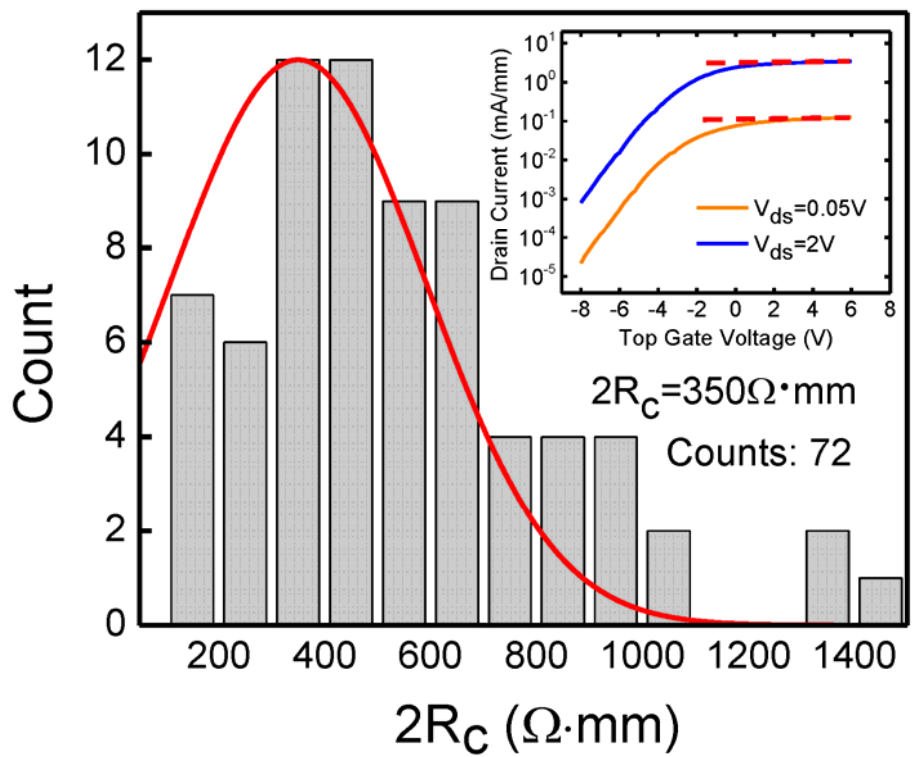

Figure 5: Intrinsic contact resistance distributions in all top-gated devices. An expectation of 350 Ω·mm is predicted by normal distribution. Inset: Typical transfer curves from a top-gated device showing early current saturation.

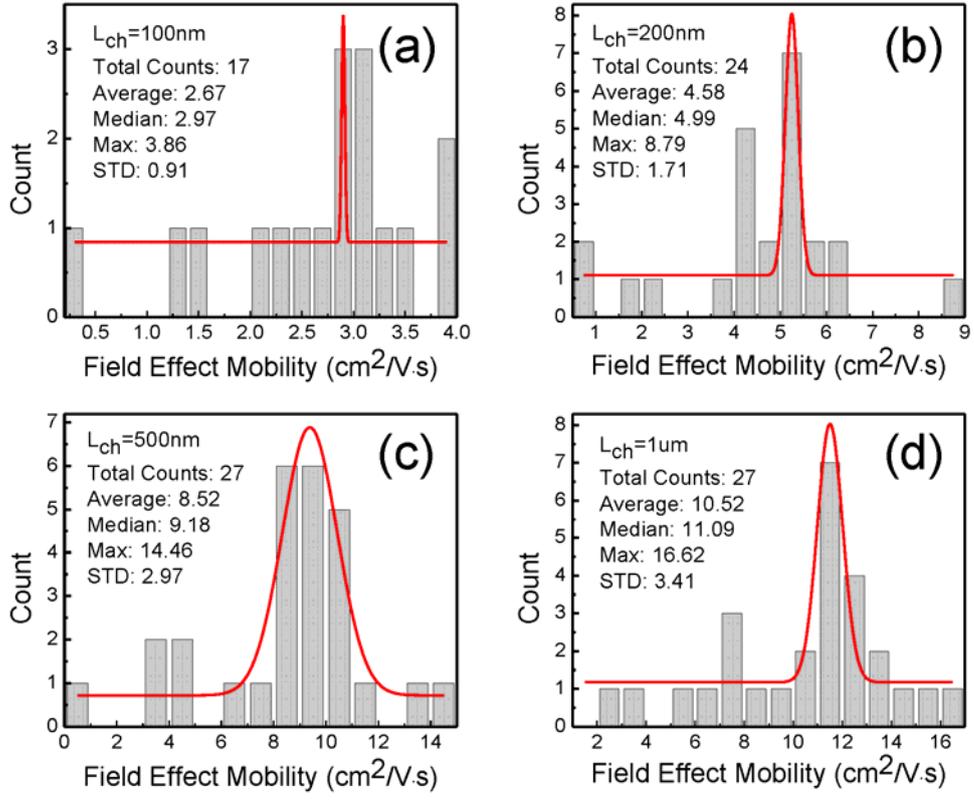

Figure 6: Extrinsic field-effect mobility distributions of (a) 100 nm, (b) 200 nm, (c) 500 nm and (d) 1 μm devices from back-gate modulation. Carrier mobilities are extracted from transconductance at low drain bias ($V_{ds}$=0.05V). Average values of 2.67±0.91, 4.58±1.71, 8.52±2.97 and 10.52±3.41 cm$^2$/V·s for 100, 200, 500 nm and 1 μm channel length are calculated.

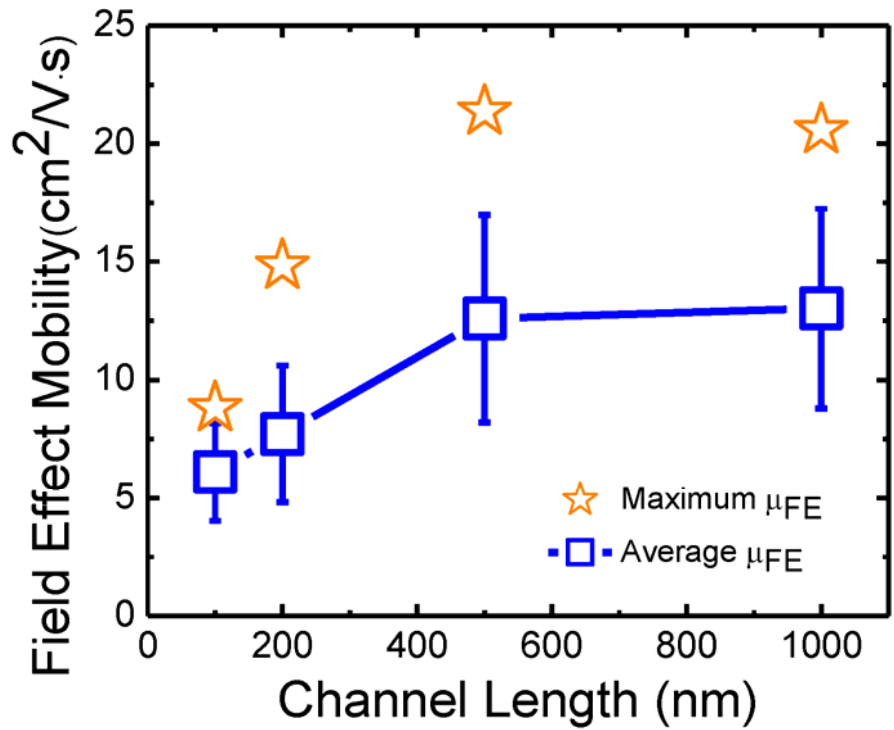

Figure 7: Average and maximum values of intrinsic carrier mobility at all channel lengths. The impact of contact resistance is subtracted from the data. The maximum values are 8.8, 14.8, 21.4 and 20.6 cm$^2$/V·s for 100, 200, 500 nm and 1 μm channel length devices.

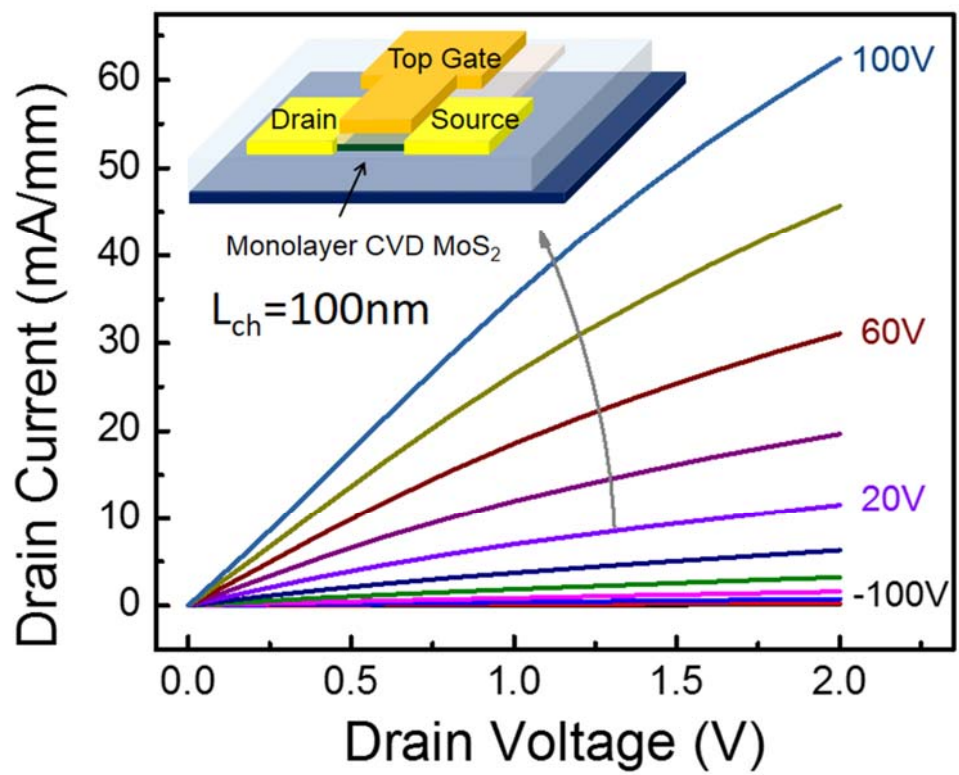

**Table of Content Graphic**

# Supporting Information

# Statistical Study of Deep Sub-micron Dual-gated Field-Effect Transistors on Monolayer CVD MoS$_2$ Films

Han Liu,[†] Mengwei Si,[†] Sina Najmaei,[‡] Adam T. Neal,[†] Yuchen Du,[†] Pulickel M. Ajayan,[‡] Jun Lou,[‡] and Peide D. Ye[†,*]

[†] School of Electrical and Computer Engineering and Birck Nanotechnology Center, Purdue University, West Lafayette, IN 47907, USA

[‡] Department of Mechanical Engineering and Materials Science, Rice University, Houston, TX 77005, USA

*Correspondance to: yep@purdue.edu

**The Role of Seeding Layers**

Figure S1

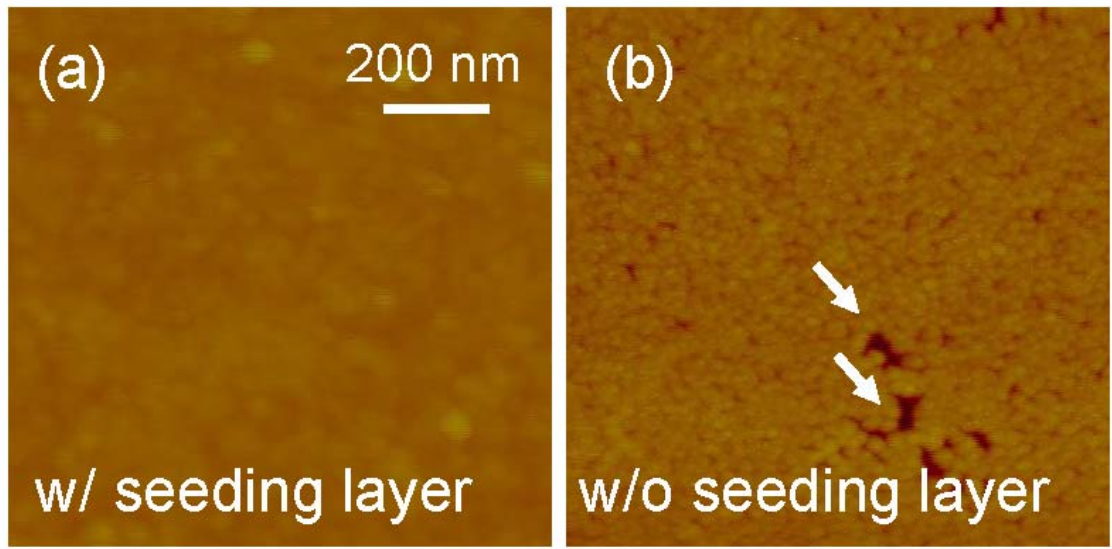

In the device fabrication process, we deposited 0.8-1 nm Al seeding layer by e-beam evaporator at a low rate of 0.1 Å per second before gate dielectric deposition. Control samples of dielectric growth without seeding layer were also fabricated for comparison to illustrate the importance and necessity of the seeding layer. These images are of $1\times 1$ μm$^2$ area and the color scale is of 40 nm. Figure S1 shows the AFM images of monolayer CVD MoS$_2$ after 15 nm Al$_2$O$_3$ growth with or without the insertion of an Al seeding layer. We clearly see from Figure S1(a) that with the insertion of the seeding layer, we achieve a uniform layer of Al$_2$O$_3$. Though the dielectric layer is not atomic flat as bare MoS$_2$ surface, they minimize gate leakage in the devices. However, with no seeding layer prior to ALD growth, as shown in Figure S1(b), though we achieve a moderate Al$_2$O$_3$ coverage in most area, we still observe pinholes, as indicated by arrows, in some parts which would deteriorate the device performance and bring down the device yield. The final results are also strongly dependent on ALD conditions. On the other hand, since the seeding layer would undoubtedly increase the interface trap density, the ALD dielectric integration on MoS$_2$ needs to be further studied and optimized.

**Output and Transfer Characteristics from Both Top-Gate Modulation and Back-Gate Modulation**

Figure S2

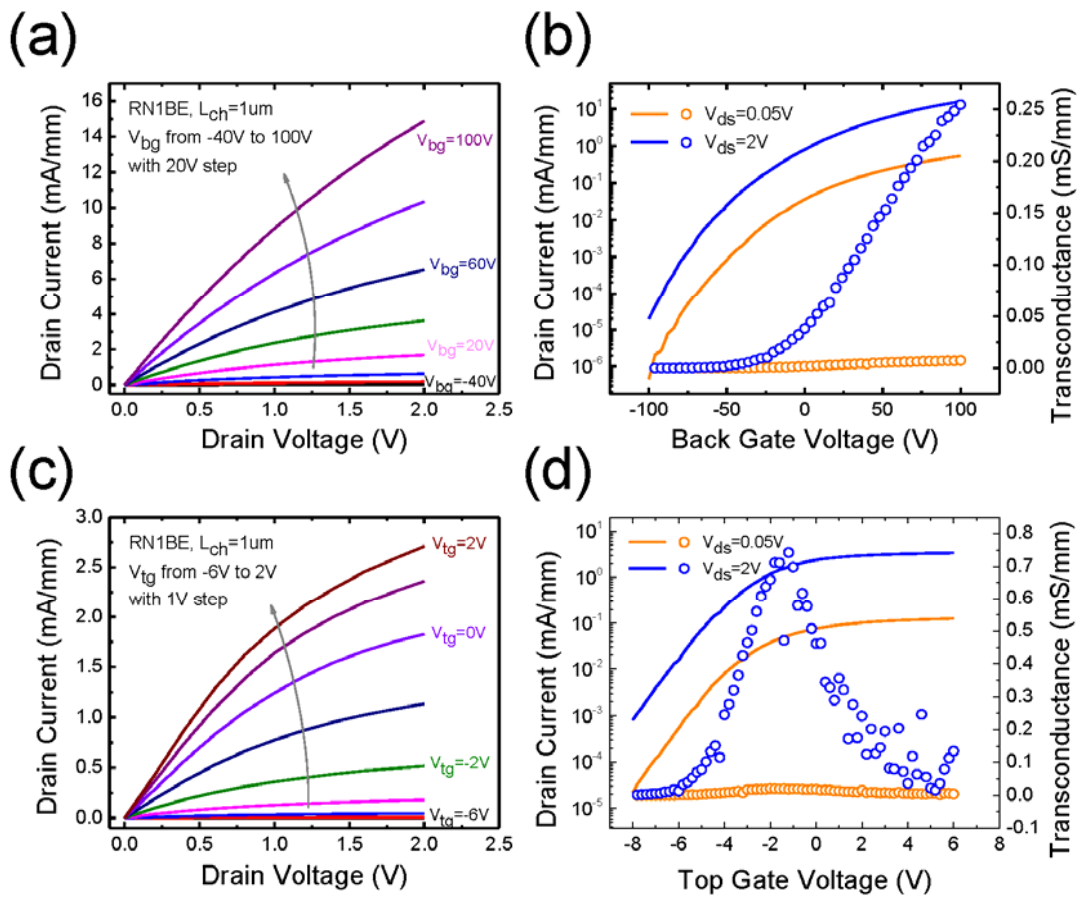

Output and transfer curves family from both top-gate and back-gate of the same device are shown in Figure S2 (a) to (d). We achieve the highest drain current of 14.95 mA/mm and 2.71 mA/mm for back-gate and top-gate modulations. The difference is mostly originated from contact resistance. The observation of linear voltage-current relationship indicates a good "ohmic" contact. Poor transconductance peak values of 0.25 mS/mm and 0.74 mS/mm are obtained for top-gate and back-gate modulations. Limitations are from thick dielectric layer (285 nm $SiO_2$), large interface trap desnity or large contact resistance. A large subthreshold swing (SS) value of 13 V/dec can be extracted from Figure S2(b). This corresponds to a mid-gap interface trap density ($D_{it}$) value of $1.6 \times 10^{13}$ $cm^{-2}eV^{-1}$ by applying $SS = \frac{kT}{q}(1 + \frac{C_{it}}{C_{ox}}) = \frac{kT}{q}(1 + \frac{qD_{it}}{C_{ox}})$, where $C_{it}$ is the interface trap capacitance, $C_{ox}$ is oxide capacitance, k is the Bolzmann constant, and T is the absolute temperature. We ascribe the large $D_{it}$ to the synthesis process, where $SiO_2$ surface was exposed in the furnace to a variety of reactive precursors at a high temperature during the CVD process.

**The Estimation of Contact Resistance**

Figure S3

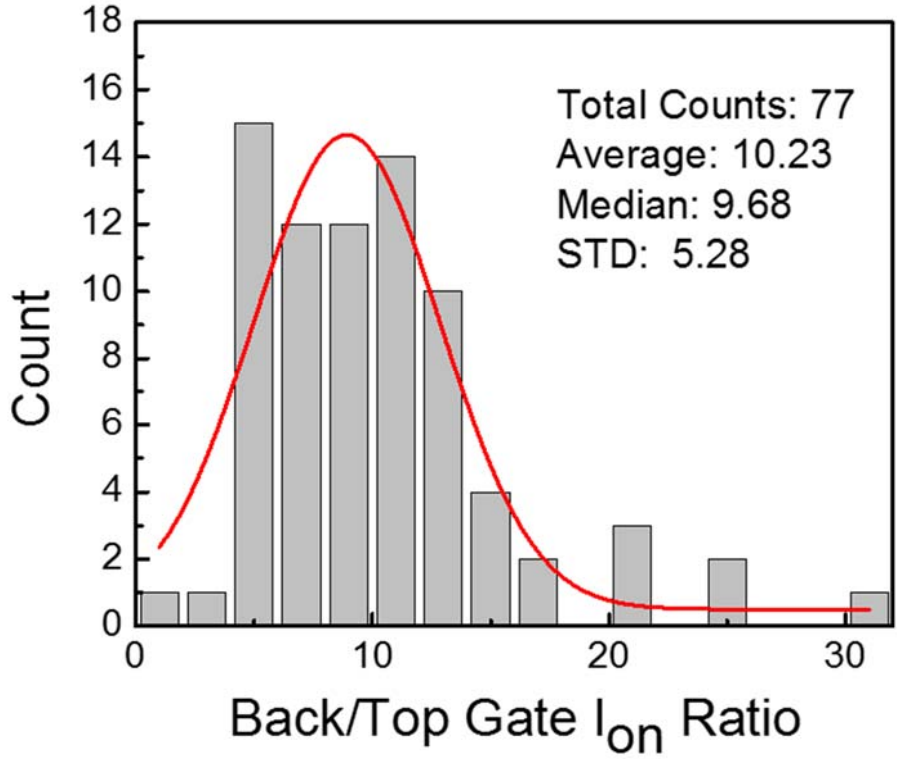

In order to extract intrinsic contact resistance, i.e. contact resistance without gate bias, we compare the total resistance at on-state of 500 nm and 1 μm channel length transistors. The ratios between maximum drain currents at 2 V drain bias from back-gate modulation over top-gate modulation are plotted in Figure S3. Both the average and median values are around 10, i.e. $R_{tot,tg} = 10 \times R_{tot,bg}$. As we have stated in the main text, the total resistance is composed by contact resistance, $2R_c$, and the channel resistance, $R_{ch}$, if we assume the channel resistance is similar from top- and back-gate modulation, this can be written as:

$$2 \times R_{c,tg} + R_{ch} = 10 \times (2R_{c,bg} + R_{ch}),$$

$$2 \times R_{c,tg} = 20 \times R_{c,bg} + 9 \times R_{ch},$$

consider even under gate bias, $R_{c,bg}$ is still considerably large, thus we will have:

$$2 \times R_{c,tg} \gg R_{ch}$$

This means, for top-gated devices, the early saturation of drain current in transfer curves (inset of Figure 5) is attributed to the large contact resistance. In another words, even channel resistance can be further decrease by gate biasing, the large contact resistance is dominant and the drain current cannot be further improved.

**The Estimation of Intrinsic Carrier Mobility**

Figure S4

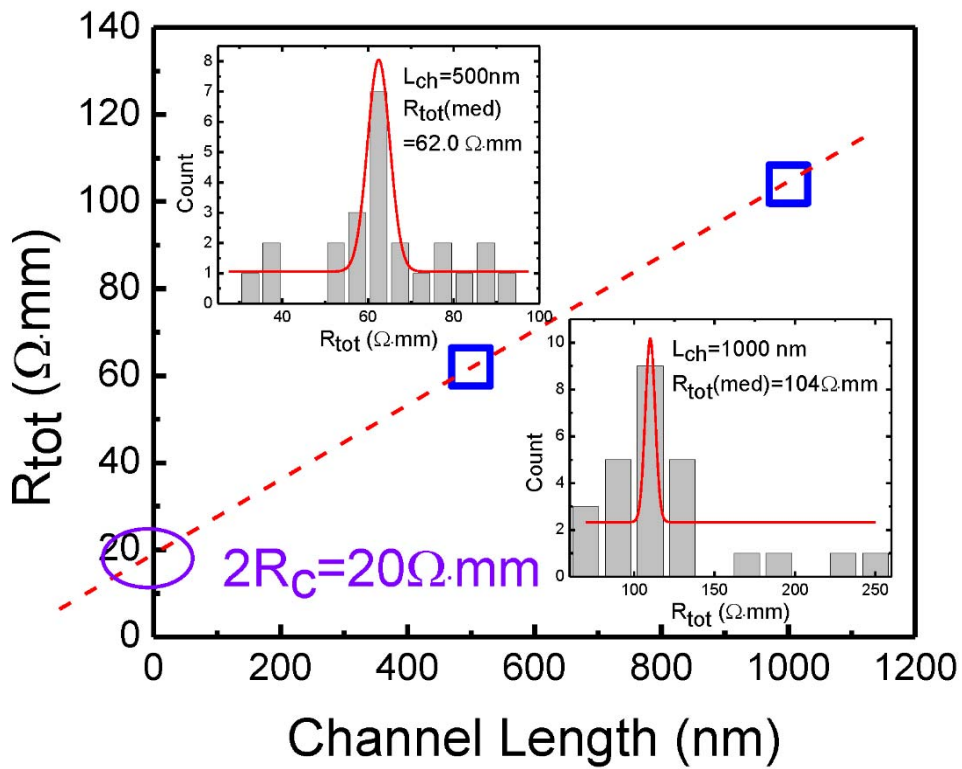

The contact resistance under gate bias is calculated by comparing the expectations of total resistance at low drain bias. We have the expectation of total channel resistance of 104 Ω·mm for 1μm channel length device, and 62 Ω·mm for 500 nm channel length device. The distribution is plotted in two insets of Figure S4. Since we have $R_{tot} = 2R_c(V_{gs}) + R_{ch}(V_g, L)$, and $R_{ch}(V_{gs}, L)|_{L=1\mu m} \approx 2R_{ch}(V_{gs}, L)|_{L=500nm}$, therefore, the contact resistance can be obtained by connecting the two total resistance in the coordinate and extend the line reversely to when channel length equals zero. 2R$_c$ would be the intercept on y-axis. From Figure S5 we get the contact resistance under 100 V back-gate bias to be 20 Ω·mm.

Figure S5

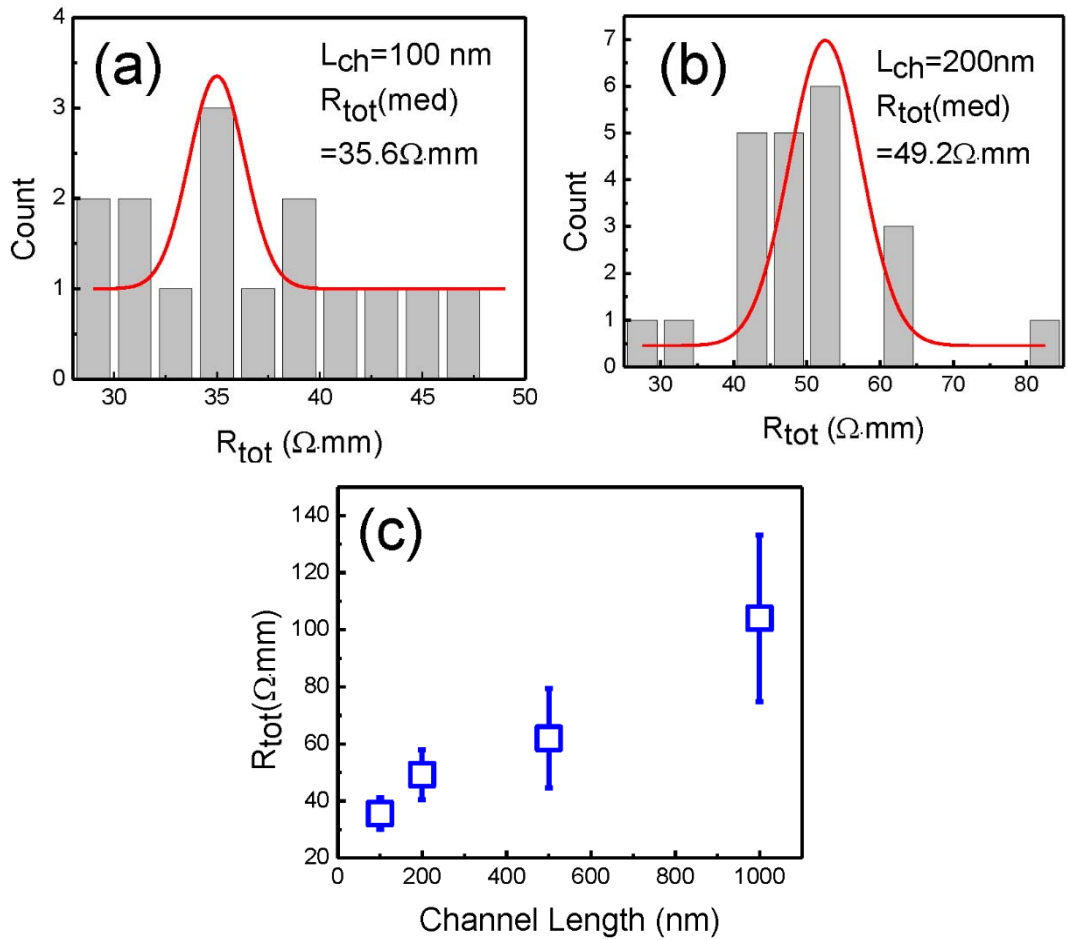

We retrieve the median values of 100 and 200 nm channel length via the same method as shown in Figure S5 (a) and (b). The total resistance versus channel length is shown Figure S5(c). Therefore we can subtract the contact resistance from the total resistance, and correct the carrier mobility by $\mu' = \mu \left( \dfrac{R_{ch}}{R_{tot}} \right)^{-1} = \mu \left( 1 - \dfrac{2R_c}{R_{tot}} \right)^{-1}$. Thus we get the multiple factors $\left( 1 - \dfrac{2R_c}{R_{tot}} \right)^{-1}$ to be 2.28, 1.68, 1.47 and 1.23 for 100, 200, 300 nm and 1 μm channel length devices to calculate the intrinsic carrier field-effect mobility.